\RequirePackage{fix-cm}
\documentclass[twocolumn]{svjour3}          
\smartqed  

\usepackage{cite}
\usepackage{amsmath,amssymb,amsfonts}
\usepackage{algorithmic}
\usepackage{graphicx}
\usepackage{textcomp}
\usepackage{xcolor}
\usepackage{graphicx}
\usepackage{subcaption}
\usepackage{csquotes}
\begin{document}

\title{On Achilles Heel of Some Optical Network Designs and Performance Comparisons 
}


\author{Dao Thanh Hai   
}


\institute{F. Author \at
	Post and Telecommunication Institute of Technology \\
	\email{haidt102@gmail.com} 
}

\date{Received: date / Accepted: date}

\maketitle

\begin{abstract}
This non-conventional paper represents the first attempt to uncover a possible vulnerability in some proposals for optical network designs and performance comparisons. While optical network designs and planning lie at the heart of achieving fiber capacity efficiency and/or operational efficiency, its combinatorial nature makes it computationally hard to reach optimal solutions for realistic scenarios. As a consequence, the well-established way that have been taken for granted by not-so-small number of research papers is that an optimization model based on mixed integer linear programming (MILP) formulation is first proposed and then due to the intractability of such combinatorial optimization model, a heuristic algorithm is offered as an approximation. The solution-quality comparison between the MILP and heuristic is then carried out on small-scale instances including topologies and traffic tests to verify the efficacy of the proposed heuristic. Next such allegedly verified heuristic are used for optical network designs of realistic scenarios. This approach may nevertheless leave a critical vulnerability as there is no guarantee that one performs well in small tests will generalize adequately for large-scale cases, a common pitfall widely referred as the peril of extrapolation and/or overfitting. Besides, it is not uncommon that in some research works, for benchmarking purpose, the comparison between a new design proposal whose performance is obtained from on one heuristic and a reference design based on another heuristic is carried out. As the consequence of lacking solution quality check, such performance comparison relied merely on heuristic solutions may be equally vulnerable, resulting to possibly unreliable conclusions. In this work, we pinpoint those issues and provide a realistic case study to highlight and demonstrate the impact of such vulnerabilities.  

\end{abstract}

\section{Introduction}
\label{intro}

Internet has become clearly the largest engineered system made by humankind with millions of end devices, telecommunication links, switches and routers connecting to each others and billions of users have been on Internet via different means. Less well-known is that the major segment of Internet infrastructure are composed of several billion kilometers of optical fibers that have been installed worldwide. Indeed, optical fibers serve as the tremendous communication channels through which nearly all voice, video, and data communications fly almost instantaneously around the globe \cite{ir4, 20years}. For many years, the capacity of an optical fiber has been viewed as being almost infinite and indeed, the single channel capacity in optical core networks has undergone leap-and-bound growth with a spectacular rise from 2.5 Gb/s in around 1990 to beyond 1 Tb/s in 2020 \cite{20years, Simmons}. Such factor of 400-fold increase in a span of roughly 30 years is thanks to convergence of remarkable advances in electronic, photonic and digital signal processing technologies \cite{hai_wiley, hai_comletter, hai_icact, hai_systems, hai_oft, hai_oft2, hai_thesis}. Nevertheless, in order to exploit such huge capacity in an efficient and sustainable manner, great emphasis has been placed on developing efficient network design algorithms \cite{hai_comcom, hai_comcom2, hai_springer2, hai_optik, hai_access, hai_csndsp, hai_icist1, hai_sigtel2}. Carefully designed optical networking algorithms have therefore emerged as one of the critical components accompanying system capacity growth. Clearly, it has been well-known that a good set of algorithms will maximize the utilization of network resources, which ultimately translates to greater capital and operation efficiency. In contrast, inadequate algorithms may result in squandering the available capacity, causing poor network performances \cite{Algorithm1, Algorithm2, hai_springer, hai_nics, hai_atc, hai_rtuwo} \\ 

When it comes to designing and operating an optical network, a central issue to be addressed is the routing and resource allocations for demands \cite{rwa1, rwa2, rmlsa, h1, h3, h4, h5, h6, h7, h8, h9}. That problem will become basically the routing and spectrum assignment (RSA) if elastic optical networks technology is employed \cite{hai_iet, hai_ps1, hai_ps2, hai_icist2, hai_sigtel1}. In the case of operating with fixed wavelength division multiplexing (WDM) technology, it is called routing and wavelength assignment (RWA) problem \cite{Simmons, rwa}. Although there have been various other variants depending on the physical layer assumptions and switching architectures of optical networks \cite{hai2021shades, hai_comletter, variant1, variant2}, RWA and RSA remain the most basic issue on which other variants are emerged. It has to be noticed that the typical procedure that have been adopted in some research papers for dealing with RWA, RSA problems and other variants follow basically the same route as following:

\begin{itemize}
	\item \textbf{Step 1:} Formulate the network design problem in the form of MILP model 
	\item \textbf{Step 2:} Complete the formulation part with a conclusion whose meaning basically is the MILP model is not well-suited for a large number of traffic demands in large-scale network topologies and therefore heuristic algorithms are needed to develop 
	\item \textbf{Step 3:} Develop one (or some) scalable heuristic algorithm 
	\item \textbf{Step 4:} Test the efficacy of one (or some) proposed heurisic by comparing its solution with that of the MILP model in a small network topology and few demands; Declare the goodness of one proposed heurisic based on the data that its solution is close to or equal to optimal ones in small-scale tests and use that victory heuristic for designing and operating the realistic (large-scale) ones 
	\item \textbf{Step 5:} Conduct a performance comparison on realistic network topologies between a new design proposal whose performance is obtained from the proposed heuristic and a reference design whose performance may be obtained from another heuristic in literature. The purpose is to report the achieved gains with new proposal compared to the reference one 
\end{itemize}

That conventional process governing a number of research papers in optical network design and performance evaluation may have some vulnerabilities. First it remains uncertain that a heuristic that produces good results on a very small network would work adequately for a realistically sized network and vice versa, that is, an heuristic with lower performance on limited small-scale tests would function inadequately in large-scale scenarios. Our observation of this vulnerability is inspired by the well-known perils of extrapolation and the overfitting issue in machine learning. The point is that it should be very cautious to generalize a model from training data to unseen data without rigorous evaluation. Second, as a consequence of uncertainty in solution quality obtained from heuristic, operating a realistic network with a heuristic may result in poor network performances and worse still, may ruin the potential benefits that a new technological solution permits. Besides, the fact that comparing two designs based on non-optimal states derived by different heuristic algorithms may undermine the conclusion as the reported figure of merit can be distorted. In this work, we pinpoint those issues and provide a realistic case study to highlight and demonstrate the impact of such vulnerabilities. For the sake of clarity, the adversarial illustration provided in this paper is just centered on solving RWA and RSA problems, yet the underlying argument is broader.   

\section{Optical Networks Design: Exact vs. Heuristic Solutions}
In this part, we present a typical process of designing and operating an optical network with exact and heuristic solutions. The case in point is a static planning problem where given a physical topology and a set of demands to be served, find the route and assign spectrum to each demand subject to a number of technological constraints so that all demands are served and the objective is to minimize the spectrum width. Two variants are considered, that is, RSA and RWA corresponding to elastic optical networks and fixed WDM ones. For convenience, only optimization model and heuristic for RSA are described here.  \\ 

\noindent \textbf{Step 1:} MILP formulation for RSA problem \\

\begin{footnotesize}
	\noindent{Given Information:}
	\begin{itemize}
		\item $G(V,E)$: Physical network topology with $|V|$ nodes and $|E|$ links. Each link $e \in E$ has its beginning node $s(e)$ and its ending node $r(e)$. 
		\item $D$: Set of traffic demand, indexed by $d$. Each demand $d$ has its origin $s(d)$ and destination $r(d)$ respectively and request $n_d$ spectrum slices
		\item $S$: Set of spectrum slices on each link, indexed by $s$. The link capacity measured in number of spectrum slices is therefore $|S|$
		\item $C^{d}$: Set of channels for demand d. Each channel $c \in C^{d}$ consists of exactly $n_d$ contiguous spectrum slices. Note that $|C^{d}|=|S|-n_d+1$
		\item $C=\bigcup\limits_{d \in D} C^{d}$: Set of all possible channels
		\item $w_{c, s}$: constant equals to 1 if the channel $c$ contains spectrum slice $s$, 0 otherwise
	\end{itemize}
\end{footnotesize}

\begin{footnotesize}
	\noindent{Variables:}
	\begin{itemize}
		\item $x_{e, c}^{d} \in \{0,1\} $: 1 if the link $e$ and channel $c$ is used for demand $d$, 0 otherwise.
		\item $\alpha_{c}^{d} \in \{0,1\} $: 1 if the channel $c$ is used for demand $d$, 0 otherwise.
		\item $\gamma_{e,s}$: 1 if the spectrum slice $s$ is used on link $e$, 0 otherwise
		\item $\delta_{s}$: 1 if the spectrum slice $s$ is used for the entire network, 0 otherwise
	\end{itemize}
\end{footnotesize}

\begin{footnotesize}
	\noindent{Objective:} 
	\begin{equation} \label{eq:obj}
	\textit{Minimize} \; \sum_{s \in S}  \delta_{s}
	\end{equation}
\end{footnotesize}

\begin{footnotesize}
	\noindent{Subject to the following constraints:}
	\begin{equation}\label{eq:c1}
	\sum_{c \in C^d} \alpha^{d}_c = 1 \; \; \forall d \in D 
	\end{equation}
	
	\begin{equation} \label{eq:c2}
	\begin{split}
	\sum_{e \in {E}: v\equiv s(e)} {x_{e, c}^{d}}-\sum_{e \in {E}: v \equiv r(e)} {x_{e, c}^{d}}= \\		
	\begin{cases} 
	\alpha_{c}^{d} &\mbox{if } v \equiv s(d) \\ 
	-\alpha_{c}^{d}  & \mbox{if } v \equiv r(d)\\
	$0$ & otherwise \\
	\end{cases}     \qquad \qquad \forall v \in V, \forall d \in D, \forall c \in C^{d} \hfill
	\end{split}
	\end{equation}

	\begin{equation} \label{eq:c3}
	{ 
		\sum_{d \in D} \sum_{c \in C^{d}} x_{e, c}^{d} \times \omega_{c, s} = \gamma_{e,s} \qquad \forall e \in E, \forall s \in S
	}
	\end{equation}
	
	\begin{equation} \label{eq:c4}
	\sum_{e \in E} \gamma_{e,s} \leq |E|\delta_{s} \qquad \forall s \in S
	\end{equation}	
	
\end{footnotesize}

The objective in Eq. \ref{eq:obj} aims at minimizing the spectrum usage measured in terms of number of spectrum slots to support all traffic demands. Constraints in Eq. \ref{eq:c1} guarantees that all demands are served by finding the right spectrum channel. The traditional flow conservation constraint of each demand is ensured in Eq.\ref{eq:c2}. The spectrum slice uniqueness is enforced by Eq. \ref{eq:c3}. Finally, the definition of using a spectrum slice in the network, that is, a spectrum slice is considered to be used if it is occupied in any link of the network, is indicated by Eq. \ref{eq:c4} \\

\noindent \textbf{Step 2:} It has been widely proven that RSA problem is NP-complete \cite{Algorithm1, rmlsa, h7} and therefore optimization solvers are not guaranteed to finish in polynomial time. A scalable heuristic solution has to be developed to operate for realistic networks. \\

\noindent \textbf{Step 3:} As a reference heuristic algorithm for RSA problem, we choose MSF (i.e., Most Slices First), a typical reference algorithm for benchmarking purpose. Details of the algorithm could be found in \cite{hai_iet, rmlsa}. MSF is a sequential heuristic algorithm for RSA where demands are ordered according to their number of requested spectrum slices and demands with highest number of slices is served first. The serving of each demand under the spectrum continuity, spectrum contiguity and non-overlapping condition is performed as follows.\\ 

We represent the spectrum availability in each link $e \in E$ by a vector $u_e$ with the size equaling to the number of spectrum slices $S$. Each element in the vector encodes the availability of the spectrum slice, that is, element $i$ equals to 1 if the slice $i$ is available and $0$ if it is occupied. 
\begin{equation}
\overline{u_e}=(u_{e1}, u_{e2},..., u_{e|S|} )
\end{equation}

For a path $p$, we can calculate the spectrum availability vector for that path by performing AND operation for all the spectrum availability vectors of links constituting the path. \\
\begin{equation}
\overline{U_p}=\cap_{e \in p}\overline{u_e}
\end{equation}

For each demand $d$ requesting $n_d$ slices, we first evaluate the spectrum availability vectors of $3$ candidate paths (i.e., 3 shortest paths) and then search the first possible placement of $n_d$ contiguous slices. The path with the lowest indexed starting slice is selected. Afterward, we update the spectrum availability of the links constituting the selected path. After serving all demands, we evaluate this solution with a fitness function as $F=max{f(\overline{u_e})}$ where $f(.)$ is the function that returns the index of the last used slice on spectrum availability vector $\overline{u_e}$. The value of $F$ is the objective value of that solution. 

\section{Mind the Perils of Extrapolation and Performance Comparison: A Case Study}
In this part, we present a numerical case study on solving the routing and spectrum assignment (RSA), and routing and wavelength assignment (RWA) with both exact and heuristic approaches. As far as this part is concerned, that the RSA problem is solved represents the choice of elastic optical networks technologies to support a given set of traffic demands while the solving of RWA problem refers to the adoption of fixed WDM technologies. For solving the MILP models, we use CPLEX and run until the optimal solution is reached. Advanced implementation techniques have been performed including warm-start and parallel computing to expedite the running time. \\

When it comes to heuristically solving the RWA problem, the metric for evaluation is the number of used wavelengths to accommodate the same set of traffics as with the RSA problem. Here we use a genetic algorithm for finding a good order of demands to be served and then the first-fit wavelength assignment is applied. Details of the algorithm could be found in \cite{hai_sigtel1, Simmons}. 

\subsection{Step 4: The Possible Trap of Extrapolation} 
We first draw a comparison on solution quality between exact approach based on solving optimally optimization model and approximation approach based on heuristic algorithms. The conventional way to do so is first to compare on small-scale instances and if the heuristic's quality is good enough, a conclusion could be made that it is ready to apply the heuristic for large-scale models. But this is where the pitfall may occur. \\ 

\begin{figure}[!ht]
	\centering
	\includegraphics[width=0.8\linewidth, height = 8cm]{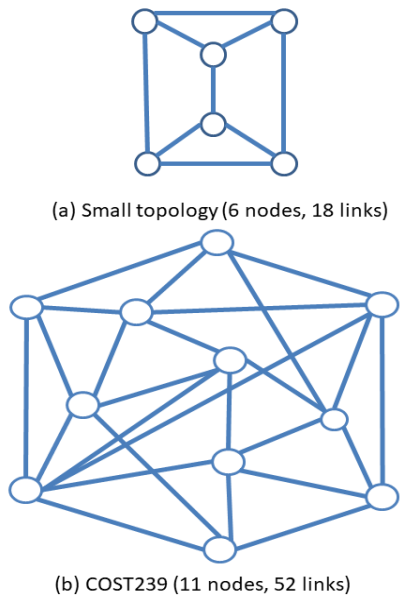}
	\caption{Network Topologies under Test}
	\label{fig:i1}
\end{figure}

In our numerical illustration, for small tests, we use a 6-node topology (Figure \ref{fig:i1}(a)) with 10 traffic matrices and each traffic matrix consists of 8 traffic demands whose number of requested spectrum slices is a random integer number between 1 and 4. This corresponds to a random demand between 25 Gbps and 100 Gbps with a step of 25 Gbps operating in an flex-grid optical network adopting QPSK modulation and spectrum slice width of 12.5 GHz. Table \ref{tab: r1} reveals the comparison of heuristic and optimal solutions for both RSA and RWA problem on small instances. Note that the number of used spectrum slices is a metric for RSA problem and each slice is 12.5 GHz width while the number of used wavelength is a metric for RWA problem and each wavelength is 50 GHz width. As can been seen from Table \ref{tab: r1}, both RSA heuristic and RWA heuristic achieve remarkable performance as its solution is optimal in majority of instances. Indeed, only at traffic instance 5, the RSA heuristic is one slice far from its optimality. Based on that observation, the conventional declaration of some research works would be that the proposed heuristic is efficient and it is therefore ready to bring that heuristic to realistic scenarios.   

\begin{table}[ht]
	\caption{Performance Comparison between Exact Solutions and Heuristic Ones}
	\label{tab: r1}
	\centering
	\begin{tabular}{|c|c|c|c|c|}
		\hline
		Traffic & Optimal & Heuristic & Optimal & Heuristic \\
		Instance & RSA & RSA & RWA & RWA \\
		\hline \hline
		1 & 4 & 4 & 2 & 2 \\
		2 & 4 & 4 & 1 & 1 \\
		3 & 4 & 4 & 1 & 1 \\
		4 & 4 & 4 & 1 & 1 \\ 
		5 & 4 & 5 & 2 & 2 \\ 
		6 & 4 & 4 & 1 & 1 \\ 
		7 & 4 & 4 & 1 & 1 \\ 
		8 & 4 & 4 & 2 & 2 \\ 
		9 & 4 & 4 & 1 & 1 \\ 
		10 & 3 & 3 & 2 & 2 \\ 
		\hline
	\end{tabular}
\end{table}

For realistic network, a COST239 topology (Figure \ref{fig:i1}(b)) is employed with 10 traffic sets whose each one is made up of 45 random demands. Each demand again requests a random number of requested spectrum slices between 1 and 4. Note that for fixed WDM network with QPSK modulation and wavelength granularity of 50 GHz, either the demand is 25 Gbps or 100 Gbps, it is all filled with one wavelength of 50 GHz. \\

\begin{figure}[!ht]
	\centering
	\includegraphics[width=\linewidth, height = 5.5cm]{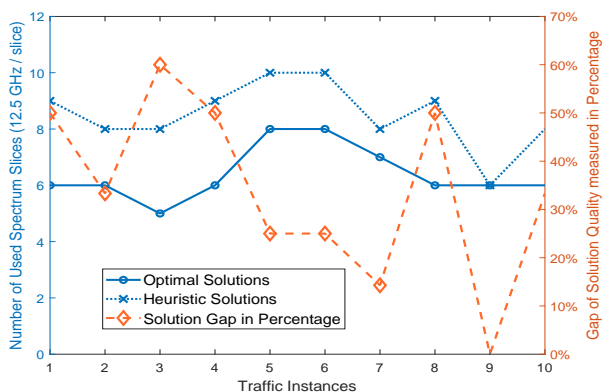}
	\caption{Comparison between Optimal and Heuristic Solutions of RWA problem on COST239 network}
	\label{fig:r1}
\end{figure}

\begin{figure}[!ht]
	\centering
	\includegraphics[width=\linewidth, height = 6cm]{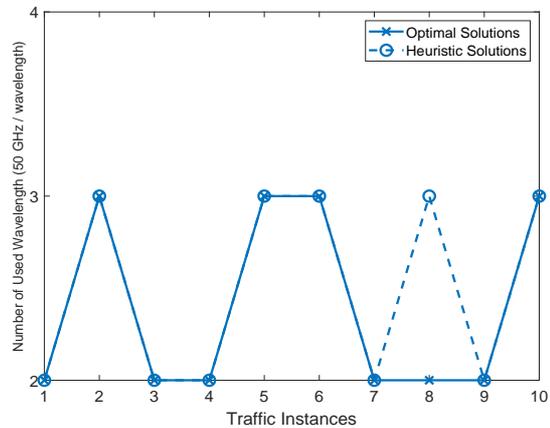}
	\caption{Comparison between Optimal and Heuristic Solutions of RSA problem on COST239 network}
	\label{fig:r2}
\end{figure}

Let us observe the results of applying such heuristic on a realistic COST 239 topology. Different from traditional research works, we provide here the optimal solutions obtained from solving the MILP formulation to double-check the heuristic solutions. Figure \ref{fig:r1} depicts the comparison of optimal RSA and heuristic one at different traffic instances. It can be seen that its heuristic solution deviate considerably from optimal values. Only 1 out of 10 cases, the heuristic solution reaches the optimal value while in the worst case, the gap can be up to $60\%$. On average, the heuristic solution has a quality gap of $34\%$ compared to optimal one. In other words, there has been a failure in the generalization of RSA heuristic from small cases to large ones. The consequence of that poor generalization is that if such heuristic is decided to use for designing and operating a real network, a substantial spectrum resources could be squandered due to the poor provisioning of algorithm. On the other hand, as shown in Figure \ref{fig:r2}, the RWA heuristic has an exceptional generalization capability as it can achieve optimality in 9 out of 10 cases and the average gap is therefore as low as $5\%$. \\

Apparently both RSA and RWA heuristic are good enough under small-scale tests and yet, they behave differently upon generalization. For these particular RSA and RWA heuristic and on the particular COST239 network, the discrepancy between optimal and heuristic ones are likely due to the flexibilities of heuristic algorithms in adapting to changing network topologies. Putting the spectrum/wavelength assignment aside as it is both relied on first-fit scheme, the RSA heuristic is based on a fixed routing mechanism of finding 3 candidate paths for each demand. On a small topology of 6-node as in Fig. 1(a), such 3 candidate paths are adequate for reaching (nearly) optimal solutions. However, when topology is changed to a more densely connected structure, such fixed 3 candidate paths routing appears to miss a large part of search space. As a result, the RSA heuristic solutions are likely deviated from the optimal ones. On the other hand, the RWA heuristic is a more advanced one permitting certain adaptation to both traffic and topology. Specifically, the Genetic Algorithm is used to find the good ordering of traffic demands for serving and for the routing, a maximum of 10 different candidate paths are allowed for each demand. Consequently, in some way, the RWA heuristic is more adaptive to a wide range of traffic and network topologies and therefore, justifies for its generalization capability.

Note that we have intentionally made use of such two carefully crafted heuristic to pinpoint a loophole, possibly resulting in Achilles heel when it comes to designing and operating optical networks with heuristic algorithms. The critical point to raise is that current practices of designing and testing heuristic algorithms have been performed without rigorous framework. In fact, it remains quite arbitrary for the majorities of works on the selection of small cases for heuristic algorithm testing and it has been lacking of useful metrics for measuring and/or predicting the effectiveness of heuristic algorithms with respect to specific features of network topologies and traffic demands.

%

%
%

\subsection{Step 5: The Possible Trap of Performance Comparison}
We now focus on the issue of performance comparison when it comes to accommodating a set of traffic demands with elastic optical networks technologies and a reference one based on WDM technologies. Basically, it is about solving the aforementioned RSA and RWA problems and comparing their solutions. The performance metric for comparison is the bandwidth which is measured by the number of used spectrum slices $\times$ 12.5 GHz for RSA cases and the number of used wavelength  $\times$ 50 GHz for RWA cases. \\

Although it is intuitive to predict that the elastic optical networks operating at a finer granular of spectrum will be spectrally more efficient than its counterpart fixed WDM networks, realizing that potential gain is a different story. The numerical results presented in this part illustrates that critical point. \\ 

\begin{figure}[!ht]
	\centering
	\includegraphics[width=\linewidth, height = 5cm]{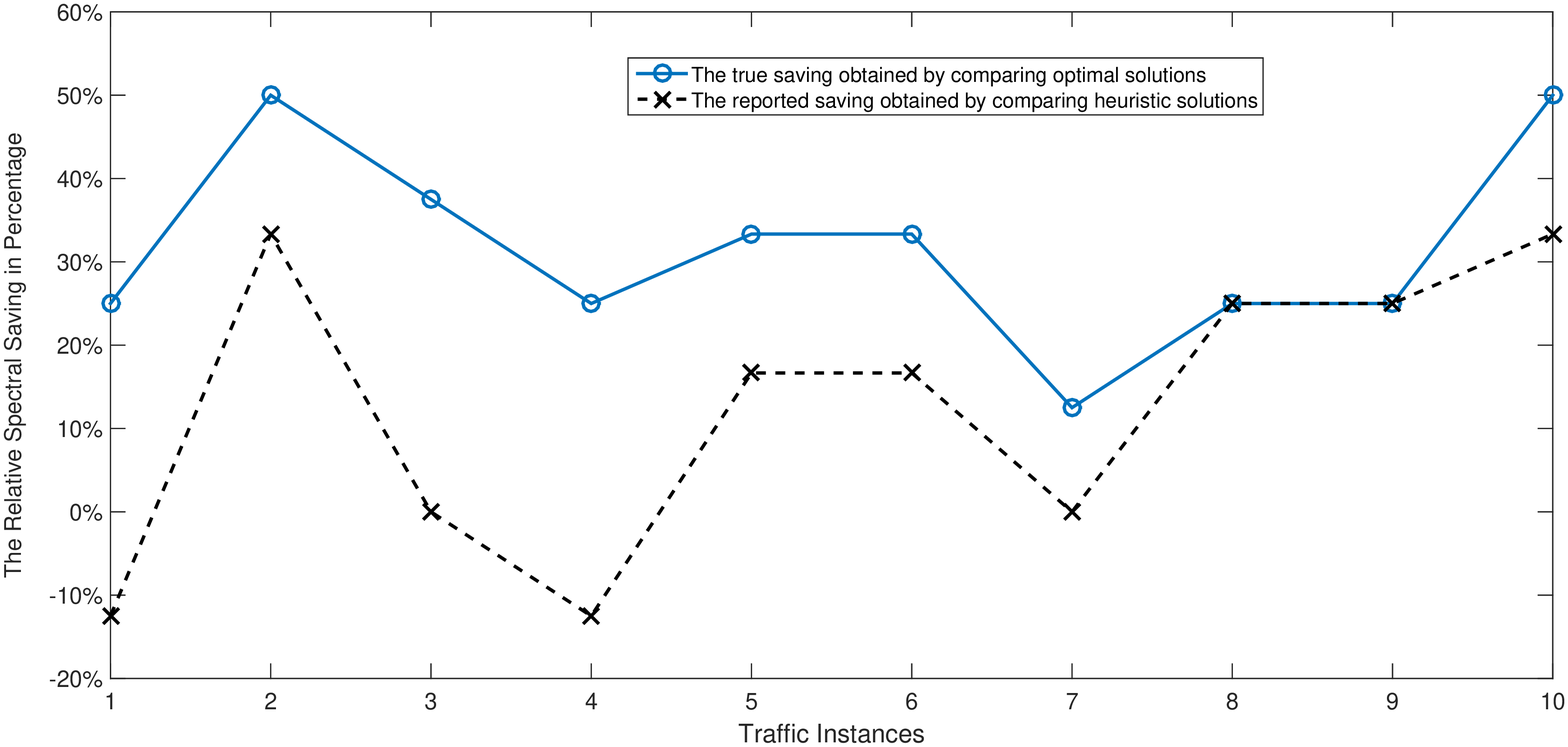}
	\caption{Relative Spectral Saving of Elastic Optical Networks compared to WDM networks}
	\label{fig:r3}
\end{figure}

Figure \ref{fig:r3} showcases the relative spectral savings when RSA (i.e., elastic optical networks is employed) is solved in comparison with solving the RWA problem (i.e., fixed WDM network is employed). Very often in some research works, only the relative spectral savings obtained from comparing RSA heuristic and RWA heuristic is reported. In an attempt to detect the vulnerability of such traditional process, we include here the (true) relative spectral savings obtained from comparing the optimal solutions of both RSA and RWA problems. This is to highlight the point that how performing with heuristic might produce distorted figure of merits. \\


As shown in Figure \ref{fig:r3}, there are two instances (i.e., traffic index 1, 4) that the relative spectral saving is negative and it means that operating an elastic optical networks with the proposed RSA heuristic results in lower spectral efficiency than operating a fixed WDM network with the RWA heuristic. This counter-intuitive result is due to the low-quality of RSA heuristic when applying to the realistic COST239 network while the RWA heuristic performs adequately well. Such comparison thus results in distorted conclusions as the relative solution quality of two heuristics are not taken into account. Furthermore, the result highlights the fact that a poorly designed heuristic algorithm may ruin all the potential benefits that a new technology can bring. Indeed, as it can be seen, operating with MSF heuristic for RSA problem may cause higher usage of spectrum resources compared to RWA counterpart in certain cases. Moreover, it is observed that how deviated the heuristic-based comparison is in reference to the true one. In this illustrative case, the reported gain concluded from relying on merely heuristic is on average roughly $20\%$ lower than the true one based on comparison of optimal solutions. \\

The aforementioned numerical illustration pinpoints to a critical vulnerability that a conclusion drawn from performance comparison of two designs based on solely heuristic algorithms may misconceive the reality. This is due to the uncertainty in quality of heuristic solutions when extending to large-scale scenarios.


\section{Summary}
The peril of extrapolation and overfitting has found its way to some proposals for optical network designs and performance comparisons and it has been addressed for the first time in this paper. The first argument was that the traditional process of testing a heuristic algorithm on small-scale cases and then extend it to large-scale ones may leave a critical vulnerability as there is always a possibility that the heuristic does not generalize well for the large-scale ones. Consequently, if that ill-generalized heuristic is used for designing and operating a realistic optical network, a substantial spectrum resources could be squandered or alternatively, the potential of a new technological and architectural proposal could not be realized. The second argument was to pinpoint a loophole in comparing two designs based on merely heuristic solutions. Due to the nature of uncertainty in heuristic solution quality, conclusions drawn from comparing performances of two designs based merely on different heuristics may be unreliable in the sense that there might be either a large gap compared to the ideal one or worse still, a false outcome (i.e., positive gain to negative one or vice versa). Numerical case studies were provided to support our arguments. \\ 

As optical network designs lie at the intersection of fiber-optic communications, operation research and computer science, more concerted effort should therefore be undertaken to enable a more rigorous evaluation and more interpretable (explainable) results to network operators. It should be pointed out that network design algorithms play a key role in achieving operational efficiency of a network and thus, more robust and carefully designed algorithms should be developed. As a quick suggestion from this paper, a reasonable practice should be to provide (lower) bounds in addition to heuristic results to have a more informed perspective. \\

Albeit still primitive, this work calls for a reconsideration in designing and evaluation of algorithms for optical network design and a better way for conducting performance comparison. Perhaps, wisdoms from machine learning realm when it comes to training/testing and combining models in the form of ensemble learning to reduce the risk of overfitting could provide some hints. Further works should be carried out to identify a framework in addressing, for example, how many small-scale cases and how diversity regarding network topologies and traffic generation are needed to test before generalization. A forward-looking perspective would also be on conceiving data-driven heuristics (e.g., topology, traffic and/or bandwidth-aware) rather than currently adopted ones with limited flexibilities.


%
\section*{Conflict of interest}
The authors declare that they have no conflict of interest.


\bibliographystyle{spmpsci_modified}      
\bibliography{ref}   

\end{document}